\documentclass[11pt]{article}
\usepackage{amssymb} 
\usepackage{amsmath} 
\usepackage{graphicx}
\usepackage{hyperref}
\begin{document}       
\title{Giant Fluctuations in Self-Propelled Particles with Age-Dependent Switching} 
\author{Shabnam Sohrabi and  Farhad H. Jafarpour\footnote{Author to whom any correspondence should be addressed.: f.jafarpour@basu.ac.ir}}
\date{Faculty of Science, Bu-Ali Sina University ,Hamedan, Iran}
\maketitle
\begin{abstract}
We investigate the transport and fluctuation properties of self-propelled particles whose motion is governed 
by an age-dependent phase-switching mechanism. The dynamics alternate between a Markovian downstream 
phase with a constant switching probability $r$ and a semi-Markovian upstream phase in which the age-dependent hazard 
probability $a/(b+c)$ decays with the internal clock $c$, representing persistent orientation. The time-averaged 
velocity, as an order parameter, shows a continuous transition at $a=1$ which separates an upstream-dominated 
ballistic regime ($a<1$) from an ergodic diffusive regime ($a>1$). Through generating-function methods and 
discrete-time moment recurrences, we derive exact expressions for the propagator and determine the long-time 
asymptotics of the mean displacement and variance. At the critical point $a=1$, the system exhibits giant fluctuations, with 
the variance scaling ballistically up to a logarithmic correction, $\mathrm{Var}(x_T) \propto T^2 / \log T$. These results 
demonstrate how slowly decaying reorientation probabilities lead to a marginal breakdown of the Central Limit 
Theorem, enabling unusually high-variance exploratory dynamics in biased environments.
\end{abstract}
\section{Introduction}\label{s1}
The transport properties of self-propelled particles in confined geometries are strongly influenced by the interplay 
between internal propulsion mechanisms and external hydrodynamic forces. In microfluidic environments, for 
example, an externally imposed flow can bias the motion of active colloids, biological swimmers, or artificial 
micromotors, leading to rich non-equilibrium behavior~\cite{Nash2010,Volpe2011,Zottl2014,Bechinger2016,Elgeti2015}.
Age-dependent switching and non-Markovian dynamics have been studied extensively in active 
matter \cite{Sadjadi2021}, anomalous diffusion \cite{Metzler2000}, and 
biological transport \cite{Korabel2018}. Related nonequilibrium transport problems also exhibit large 
current fluctuations in driven diffusive systems \cite{Bodineau2004}. 

Our work contributes to this broader picture by providing an exact analytical treatment of the fluctuation 
scaling at the marginal transition point.
The model studied in this work consists of a single self-propelled particle moving inside a tube that contains a steady background 
fluid flow.  It was initially introduced in \cite{SJ2026}, where its dynamics were analyzed using 
Large Deviation Theory (LDT)~\cite{Touchette2009,Touchette2018}. While that study established the existence 
of a dynamical phase transition at a critical threshold within trajectory ensembles, the exact scaling behavior of the transport 
moments remained uncharacterized. In the present work, we shift the focus to the underlying parameter space to 
provide an exact analytical derivation of the mean displacement and variance. This approach allows us to quantify
the ``giant fluctuations'' that characterize the critical point. We demonstrate that while the transition was 
originally identified as a non-analyticity in the large deviation rate function, its physical manifestation in the bulk statistics 
is a qualitative jump in the variance scaling---from standard diffusion to a quasi-ballistic regime---providing new insights 
into the non-ergodic nature of self-propelled particles under age-dependent switching protocols.

The particle possesses an internal engine capable of propelling it into either phase 1 which will be called \emph{downstream}, aligned with 
the flow, or phase 2 which will be called \emph{upstream}, against the flow. These two propulsion orientations generate distinct dynamical phases 
characterized by different stability properties and reversal statistics.
When the particle propels downstream, the internal engine is aligned with the ambient fluid flow, and both forces cooperate 
to push the particle forward. This mode is mechanically stable: downstream propulsion persists until the internal motor 
spontaneously reverses direction with a constant probability per step $r$. As a consequence, downstream run lengths 
are memoryless and geometrically distributed. Physically, this reflects a situation in which motion assisted by the flow 
does not accumulate mechanical stress and exhibits no long-term memory effects.
In contrast, upstream propulsion requires the particle to push against the background flow. Hydrodynamic drag opposes 
the motion, creating a complex interaction between the particle's orientation and the flow streamlines. 
To capture the resulting semi-Markovian dynamics, we introduce an internal ``reinforcement clock'' $c$ that measures the time elapsed 
since the current upstream visit began. As the particle continues to swim against the flow, the probability of a reversal 
from upstream to downstream is governed by the age-dependent hazard probability $a/(b+c)$, where $a$ and $b$ are fixed 
parameters. Crucially, the switching probability decreases with the run duration, implying that the longer a particle manages 
to maintain its upstream orientation, the more stable its trajectory becomes. This leads to a history-dependent distribution 
of upstream run lengths characterized by a heavy power-law tail. Physically, such an age-dependent reversal probability is 
motivated by experiments on intracellular transport, where molecular motors exhibit a detachment rate that decays inversely 
with the time since attachment \cite{Korabel2018}. This suggests that persistent motion against an opposing force becomes 
progressively more stable over time, a feature we capture in our minimal discrete-time framework.

The dynamics thus consist of an alternation between a stable, flow-assisted downstream phase with a constant reversal 
hazard and an adaptive, flow-opposed upstream phase with an age-dependent hazard probability. This competition produces
 nontrivial large-scale transport behavior. In particular, the statistics of the upstream run lengths may acquire heavy tails 
 whose moments diverge, suppressing downstream drift and potentially generating a net upstream motion. As $a$ (as a 
 control parameter) is tuned, the system undergoes a continuous transition in which the mean upstream run time diverges at a 
 critical point, leading to a second-order non-equilibrium phase transition at $a=1$ where the system exhibits ``giant fluctuations'' 
 that is the variance scales ballistically with a logarithmic correction.

The model considered here provides a minimal framework for studying the interplay between active 
propulsion and history-dependent stability. Despite its simplicity, it captures a 
nontrivial feedback mechanism between motility and internal state dynamics, exhibiting a sharp macroscopic 
transition in fluctuations driven entirely by the temporal structure of the upstream navigational process. 
Our analysis identifies three distinct transport regimes governed by the hazard exponent $a$. In the
 subcritical regime ($a < 1$), the walker is effectively trapped in the upstream phase due to the 
 divergence of the mean dwell time, resulting in a ballistic drift and fluctuations remain 
 asymptotically linear in time, despite the heavy-tailed run lengths.
At the critical point ($a = 1$), the system manifests a marginal heavy-tailed behavior. While the mean 
displacement remains ballistic, the fluctuations exhibit an anomalous scaling law, $\mathrm{Var}(x_T) \sim T^2 / \log T$, a 
result of the logarithmic singularity in the upstream kernels. In the supercritical regime ($a > 1$), the process 
admits a stationary phase mixture. We derive an exact expression for the effective drift and demonstrate that the 
fluctuations around the ballistic front scale linearly with time.

This paper is organized as follows. In Sec.~\ref{s2} we introduce the stochastic
switching model. Section~\ref{s3} presents the exact master equations governing the two
dynamical phases. In Sec.~\ref{s4} we develop the generating--function formalism and
derive closed expressions for the propagators. In Sec.~\ref{s5} we review the 
formalism used to compute the moments. Section~\ref{s6} analyzes the
long-time behavior of the moments in the three regimes $a<1$, $a=1$, and $a>1$.
Section~\ref{s7} presents numerical simulations supporting the analytical results. Finally, Sec.~\ref{s8}
summarizes our findings and outlines possible extensions.
\section{Model Definition}\label{s2}
We consider a discrete-time random walk $x_t\in\mathbb{Z}$, with an internal ``phase'' variable $\sigma_t\in\{+1,-1\}$, interpreted as
$\sigma_t=+1$ for downstream phase and $\sigma_t=-1$ as upstream phase.
At each time step $t\to t+1$ a hop $Y_t\in\{+1,-1\}$ is taken, with bias controlled by $\sigma_t$ so that if $\sigma_t=+1$: 
$$
    Y_t =
    \begin{cases}
    +1 & \text{with prob.\ }p,\\[4pt]
    -1 & \text{with prob.\ }1-p,
    \end{cases}
    \quad\Rightarrow\quad \mathbb{E}[Y_t\mid\sigma_t=+1] = (2p-1)\equiv v,
$$
and if $\sigma_t=-1$:
$$
    Y_t =
    \begin{cases}
    -1 & \text{with prob.\ }p,\\[4pt]
    +1 & \text{with prob.\ }1-p,
    \end{cases}
    \quad\Rightarrow\quad \mathbb{E}[Y_t\mid\sigma_t=-1] = -(2p-1)=-v.
$$
 So in general
$$
  \mathbb{E}[Y_t\mid\sigma_t]=v\,\sigma_t,\quad v:=2p-1.
$$
The position is updated by:
$$
  x_{t+1}=x_t + Y_t.
$$
The phase $\sigma_t$ evolves according to age-dependent upstream hazard and constant downstream hazard.
If $\sigma_t=+1$ at time $t$, we call the process downstream. The phase switches to upstream with a constant hazard $r$:
$$
    \mathbb{P}(\sigma_{t+1}=-1\mid \sigma_t=+1)=r,\qquad
    \mathbb{P}(\sigma_{t+1}=+1\mid \sigma_t=+1)=1-r.
$$
If $\sigma_t=-1$, the phase is upstream and carries an integer ``age'' $c_t\ge1$ (number of consecutive upstream steps so far). 
The hazard to leave upstream at age $c$ is given by:
$$
    h_c = \frac{a}{b+c},\quad 0 \le a \le b+1,\;b \ge 0.
$$
Thus
$$
    \mathbb{P}(\sigma_{t+1}=+1\mid \sigma_t=-1, c_t=c)=h_c,
$$
    and if the process remains upstream, the age increases:
$$
    \mathbb{P}(\sigma_{t+1}=-1, c_{t+1}=c+1\mid \sigma_t=-1, c_t=c)=1-h_c.
$$
A new upstream run always starts at age $c=1$ immediately after a downstream-to-upstream switch.
We are interested in the long-time asymptotics of:
$$
\langle x_T\rangle,\qquad
\langle x_T^2\rangle,\qquad
\operatorname{Var}(x_T)=\langle x_T^2\rangle-\langle x_T\rangle^2,
\quad T\to\infty.
$$
The joint process \((x_t,\sigma_t,c_t)\) is Markovian, with independent parameters \(r\), \(p\), and \(h_c\).
In the following section we will use the master equation approach to calculate these asymptotics. 
\section{Master Equations}\label{s3}
Let use define $P_1(x,t)$ as the probability that at time $t$ the walker is at position $x$ and in downstream phase $\sigma_t=+1$ and
$P_2(x,c,t)$ as the probability that at time $t$ the walker is at $x$, in upstream phase $\sigma_t=-1$, and has upstream age $c\ge1$.
Then the total probability to be at $x$ at time $t$ is given by:
$$
P(x,t) = P_1(x,t) + \sum_{c\ge1} P_2(x,c,t).
$$
We can now write the discrete master equations, encoding the hopping and switching described above as:
\begin{eqnarray}
P_1(x,t+1) &=& (1-r)\big[\,p P_1(x-1,t) + (1-p)P_1(x+1,t)\,\big] \label{me1}\\
\quad &+& \sum_{c\ge1}\Big[(1-p)h_c P_2(x-1,c,t) + p h_c P_2(x+1,c,t)\Big], \nonumber\\
P_2(x,1,t+1) &=& r\big[\,p P_1(x-1,t) + (1-p)P_1(x+1,t)\,\big] \label{me2}
\end{eqnarray}
and for $c\ge 2$:
\begin{equation}
P_2(x,c,t+1) = (1-h_{c-1})\big[\, (1-p)P_2(x-1,c-1,t) + p P_2(x+1,c-1,t)\,\big]. \label{me3}
\end{equation}
Equations (\ref{me1})-(\ref{me3}) fully define the evolution.
\section{Generating Functions}\label{s4}
We now transform the master equations (\ref{me1})-(\ref{me3})  using spatial Fourier transform and temporal $z$-transform. 
This will give us a compact expression for the propagator in $(k,z)$ space, from which we can extract moments.
\subsection{Fourier transform in space}
Let us introduce:
$$
\hat P_1(k,t) := \sum_{x\in\mathbb{Z}} e^{ikx}P_1(x,t)\;\; \text {and} \;\;
\hat P_2(k,c,t) := \sum_{x\in\mathbb{Z}} e^{ikx}P_2(x,c,t).
$$
We also define the step kernels as follows: for downstream ($\sigma=+1$):
$$
 H_+(k) := p e^{ik} + (1-p)e^{-ik} = \cos k + i(2p-1)\sin k.
$$
 and for upstream ($\sigma=-1$), the probabilities for +1 and $-1$ are swapped:
$$
H_-(k) := (1-p)e^{ik} + p e^{-ik} = \cos k - i(2p-1)\sin k.
$$
These encode the single-step hopping behavior in Fourier space.
Applying the spatial Fourier transform to (\ref{me1})-(\ref{me3}), we obtain:
\begin{eqnarray}
\hat P_1(k,t+1) &=& (1-r) H_+(k)\hat P_1(k,t) + H_-(k) \sum_{c\ge1} h_c \hat P_2(k,c,t), \label{ft1}\\
\hat P_2(k,1,t+1) &=& r H_+(k)\hat P_1(k,t), \label{ft2} \\
\hat P_2(k,c,t+1) &=& (1-h_{c-1})H_-(k)\hat P_2(k,c-1,t) \; \text{for} \; c\ge2. \label{ft3}
\end{eqnarray}
\subsection{$z$-transform in time}
We define the $z$-transforms in time as:
$$
\tilde  P_1(k,z) := \sum_{t\ge0} z^t \hat P_1(k,t),\qquad
\tilde P_2(k,c,z) := \sum_{t\ge0} z^t \hat P_2(k,c,t).
$$
Assume we start at $t=0$ in the downstream phase at the origin:
$$
P_1(x,0) = \delta_{x,0},\qquad P_2(x,c,0)=0,
$$
therefore:
$$
\hat  P_1(k,0)=1,\qquad \hat  P_2(k,c,0)=0.
$$
We now apply the $z$-transform to (\ref{ft1})-(\ref{ft3}). For convenience we define:
\begin{equation}
S(k,z):=\sum_{c\ge1} h_c \tilde  P_2(k,c,z). \label{S}
\end{equation}
After some straightforward simplifications we find:
\begin{eqnarray}
\tilde P_1(k,z) &=& \frac{1 + z H_-(k)S(k,z)}{1 -z (1-r) H_+(k)}, \label{zt1}\\
\tilde P_2(k,1,z) &=& r z H_+(k) \tilde P_1(k,z). \label{zt2}
\end{eqnarray}
For $c\ge2$ we also find:
\begin{equation}
\tilde P_2(k,c,z) = z H_-(k)(1-h_{c-1}) \tilde  P_2(k,c-1,z)\; \text{for} \;  c\ge2. \label{zt3}
\end{equation}
We will shortly see that the entire upstream sector is expressed in terms of $\tilde  P_1(k,z)$.
We now define two upstream ``age kernels'' that encode the effect of the age-dependent hazard in Fourier--$z$ space.
Let us define:
\begin{eqnarray}
\Lambda(k,z) &:=& z H_-(k), \\
\Phi(\Lambda) &:=& \sum_{c\ge1} \Lambda^c \prod_{j=1}^{c-1}(1-h_j), \\
\Psi(\Lambda) &:=& \sum_{c\ge1} h_c \Lambda^c \prod_{j=1}^{c-1}(1-h_j).
\end{eqnarray}
The domains are \(k\in[0,2\pi]\) (with periodic extension) and \(z\in\mathbb{C}\) with the physical region \(z\in[0,1)\) for 
convergence; \(\Lambda(k,z)\in\mathbb{C}\). From (\ref{zt2}) and (\ref{zt3}) we obtain:
\begin{equation}
\tilde {P}_2(k, c, z) = r z H_+(k) \tilde {P}_1(k, z) \Lambda^{c-1} \Big[\prod_{j=1}^{c-1} (1 - h_j) \Big] \; \text{for} \;  c\ge1\label{zt4}
\end{equation}
where for $c=1$ the empty product is $1$, consistent with (\ref{zt2}). 
The total probability of being in the upstream phase, $\tilde {P}_{\text{up}}(k, z)$, is the sum over all ages $c \ge 1$:
\begin{equation}
\tilde {P}_{\text{up}}(k, z) = \sum_{c \ge 1} \tilde {P}_2(k, c, z) = \frac{r z H_+(k)}{\Lambda} \tilde {P}_1(k, z) \Phi(\Lambda)=
r\frac{H_+(k)}{H_-(k)} \Phi(\Lambda) \tilde {P}_1(k, z) 
\end{equation}
where $\Phi(\Lambda)$ is the upstream occupancy kernel.
Using (\ref{S}) and (\ref{zt4}) we can write:
$$
\begin{aligned}
S(k,z) &= \sum_{c\ge1} h_c \tilde  P_2(k,c,z)\\
&=\frac{r z H_+(k)}{\Lambda} \tilde {P}_1(k, z) \Psi(\Lambda)\\
&=r \frac{H_+(k)}{H_-(k)} \tilde {P}_1(k, z) \Psi(\Lambda).
\end{aligned}
$$
Substituting $S(k, z)$ into (\ref{zt1}) and solving for $\tilde {P}_1(k, z)$ yields:
\begin{equation}
\tilde {P}_1(k, z) = \frac{1}{1 - z H_+(k) \left[ (1-r) + r \Psi(\Lambda) \right]}.
\end{equation}
This expression represents the probability of being downstream, where the denominator effectively describes the cycle of staying 
downstream or switching and returning.
The total propagator $\tilde{P}(k, z)$ is the sum of the probabilities of being in either the downstream or upstream states:
\begin{equation}
\tilde{P}(k, z) = \tilde {P}_1(k, z) + \tilde {P}_{\text{up}}(k, z) =\tilde {P}_1(k, z) \left[ 1 +r \frac{H_+(k)}{H_-(k)} \Phi(\Lambda) \right].
\end{equation}
By simplifying the algebraic coupling between the entry flux and the phase occupancy, the expression reduces to the final exact form:
\begin{equation}\label{finalP}
\tilde{P}(k, z) = \frac{1 + r \frac{H_+(k)}{H_-(k)} \Phi(\Lambda)}{1 - z H_+(k) \left[ (1-r) + r \Psi(\Lambda) \right]}.
\end{equation}
This formula packages all information regarding the age-dependent upstream hazard into the 
kernels $\Phi$ and $\Psi$, providing the starting point for asymptotic moment extraction.
\section{Moments in Fourier--$z$ Space}\label{s5}
We now extract $\langle x_T\rangle$ and $\langle x_T^2\rangle$ from $\tilde P(k,z)$.
By definition we have:
$$
\tilde P(k,z) = \sum_{T\ge0} z^T \,\mathbb{E}[e^{ik x_T}],
$$
therefore:
$$
\mathbb{E}[e^{ikx_T}] = [z^T]\tilde P(k,z),
$$
and finally:
$$
\langle x_T\rangle = \left.\frac{1}{i}\frac{\partial}{\partial k}\mathbb{E}[e^{ikx_T}]\right|_{k=0},
\quad
\langle x_T^2\rangle = -\left.\frac{\partial^2}{\partial k^2}\mathbb{E}[e^{ikx_T}]\right|_{k=0}.
$$
At the level of the generating function in $T$,
\begin{eqnarray}
M_1(z):=\sum_{T\ge0} z^T \langle x_T\rangle
= \left.\frac{1}{i}\frac{\partial}{\partial k}\tilde P(k,z)\right|_{k=0}, \\
M_2(z):=\sum_{T\ge0} z^T \langle x_T^2\rangle
= -\left.\frac{\partial^2}{\partial k^2}\tilde P(k,z)\right|_{k=0}.
\end{eqnarray}
Asymptotics as $T\to\infty$ correspond to the behavior of $M_1(z)$, $M_2(z)$ as $z\to1^-$. We therefore analyze the singular 
structure of $\tilde P(k,z)$ near $(k,z)=(0,1)$ and use Tauberian theorems \cite{Korevaar2004}.
While higher cumulants can be extracted from the same generating function by further derivatives with respect to \(k\), the variance 
alone suffices to establish the anomalous scaling at criticality, since it directly quantifies the width of the distribution and captures the 
marginal breakdown of the Central Limit Theorem.
\section{Singular Structure of the Upstream Kernels}\label{s6}
To determine the long-time asymptotic behavior of the random walk, we must characterize the functional forms of the 
kernels $\Phi(\Lambda)$ and $\Psi(\Lambda)$ in the vicinity of the critical point $\Lambda = 1$. This limit corresponds to the 
large-age regime of the upstream phase and governs the late-time dynamics of the propagator.
Recall the age-dependent hazard probability and the associated survival probability:
$$
h_c = \frac{a}{b+c}, \quad S(c) := \prod_{j=1}^c (1-h_j).
$$
Asymptotically, the survival probability can be expressed in terms of Gamma functions, yielding a power-law decay of the form:
$$
S(c) \sim C(a,b) c^{-a}, \quad \text{where} \quad C(a,b) = \frac{\Gamma(b+1)}{\Gamma(b+1-a)}.
$$
More specifically, for large $c$, the product appearing in the kernel definitions satisfies:
$$
\prod_{j=1}^{c-1}(1-h_j) = \frac{S(c-1)}{S(0)} \sim \tilde C(a,b) c^{-a},
$$
for a positive constant $\tilde C(a,b)$. Consequently, the upstream kernels admit the following asymptotic representations as polylogarithmic-type series:
\begin{eqnarray}
\Phi(\Lambda) &=& \sum_{c \ge 1} \Lambda^c \prod_{j=1}^{c-1}(1-h_j) \sim \tilde C(a,b) \sum_{c \ge 1} c^{-a} \Lambda^c, \\ 
\Psi(\Lambda) &=& \sum_{c \ge 1} h_c \Lambda^c \prod_{j=1}^{c-1}(1-h_j) \sim \tilde C(a,b) \sum_{c \ge 1} \frac{a}{b+c} c^{-a} \Lambda^c. 
\end{eqnarray}
The analytic behavior of these series as $\Lambda \to 1^-$ is critically sensitive to the exponent $a$:
\begin{itemize}
    \item \textbf{Subcritical regime ($a < 1$):} Both kernels exhibit a strong power-law divergence, indicating an infinite mean dwell time in the upstream phase.
    \item \textbf{Critical regime ($a = 1$):} The kernels manifest marginal divergence characterized by logarithmic singularities.
    \item \textbf{Supercritical regime ($a > 1$):} Both series converge at $\Lambda = 1$, implying a finite mean upstream dwell time. 
\end{itemize}
The leading-order singular forms required for the subsequent analysis are summarized as follows: for $a < 1$ the kernels diverge sufficiently to induce 
upstream localization, thereby simplifying the asymptotic dynamics of the walker. Hereafter, we refer to this regime as upstream 
localization, meaning localization in the internal dynamical state (persistent residence in the upstream phase) rather than spatial 
localization of the particle, whose motion remains ballistic.
For $a = 1$ a marginal case arises, dominated by $\log(1-\Lambda)$-type singularities that introduce logarithmic corrections to the transport coefficients.
Finally for $a > 1$ the system enters a finite-mean regime. For the specific interval $1 < a < 1+b$ relevant to this model, the defining physical 
feature is that the upstream mean sojourn time is finite.
In the following sections, we analyze each parameter regime by examining the singular structure of the 
denominator of $\tilde{P}(k, z)$ as $z \to 1^-$ and $k \to 0$.
\subsection{Regime $a<1$: Upstream Localization (Subcritical)}
For $a<1$, the upstream dwell mean diverges. The age kernels $\Phi(\Lambda)$, $\Psi(\Lambda)$ diverge as $\Lambda\to1^-$. Physically, the 
process overwhelmingly remains upstream for long times, so the phase $\sigma_t$ is almost always $-1$.
In the master-equation/generating-function description, this manifests as follows: the main contribution to the denominator of 
$\tilde P(k,z)$ comes from the upstream side; the effective long‑time dynamics converges to a single-phase biased walk with drift $-v$.

More concretely, at leading order in time, the walker is almost always upstream, so the effective motion is just a biased random walk
and hence:
$$
\mathbb{E}[Y_t]\approx -v = -(2p-1),\qquad
\mathrm{Var}(Y_t)\approx 4p(1-p).
$$
Therefore, from standard properties of biased nearest-neighbor walks:
\begin{eqnarray}
\langle x_T\rangle &\sim& -v\,T = -(2p-1)T, \nonumber \\
\langle x_T^2\rangle &\sim& v^2 T^2 + 4p(1-p)T, \\
\mathrm{Var}(x_T) &=& \langle x_T^2\rangle-\langle x_T\rangle^2 \sim 4p(1-p)T.\nonumber
\end{eqnarray}
These results can be justified rigorously by examining the pole of $\tilde P(k,z)$ near $(k,z)=(0,1)$: the dominant singularity 
corresponds to a simple pole in $z$ whose position in $k$ encodes the ballistic drift $-v$, and the second derivative in $k$ 
around that pole yields the diffusive $T$ term.
\subsection{Regime $a=1$: Critical (Marginal Heavy Tail)}
At the critical value $a=1$, the upstream dwell-time distribution has an
infinite mean but is \emph{marginally} heavy-tailed: the survival probability
decays as $S(c) \sim b/(b+c)$, which corresponds to a borderline
$1/c$ tail.  This produces logarithmic corrections in the generating
functions.  Using $h_c = 1/(b+c)$ and $S(c)=b/(b+c)$, the upstream age kernels
can be written as:
$$
\Phi(\Lambda) 
= \sum_{c\ge1} \Lambda^c S(c)
= b \sum_{c\ge1} \frac{\Lambda^c}{b+c},
\qquad
\Psi(\Lambda)
= \sum_{c\ge1} h_c \Lambda^c S(c)
= b \sum_{c\ge1} \frac{\Lambda^c}{(b+c)^2}.
$$
Standard asymptotics for Dirichlet-type series imply, as $\Lambda\to 1^-$,
\begin{align}
\Phi(\Lambda) &\sim -\,B_1(b)\,\log(1-\Lambda), \\
\Psi(\Lambda) &\sim \Psi(1) - B_2(b)\,(1-\Lambda)\log(1-\Lambda),
\end{align}
with positive constants $B_1(b)=b$ and $B_2(b)=b$, and where $\Psi(1)$ is
finite.  Thus $\Phi$ diverges logarithmically, whereas $\Psi$ remains finite
but possesses a logarithmically divergent derivative, which is the
source of the marginal corrections. Let:
$$
D(k,z) := 1 - z H_+(k)\bigl[(1-r)+ r\,\Psi(\Lambda)\bigr],\qquad 
\Lambda = z H_-(k).
$$
We expand near $(k,z)=(0,1)$, writing $\epsilon := 1-z$ and using:
$$
H_+(k) = 1 + i v k - \tfrac12(1-v^2)k^2 + O(k^3),\qquad
H_-(k) = 1 - i v k - \tfrac12(1-v^2)k^2 + O(k^3),
$$
where $v=2p-1$.  Then:
$$
1-\Lambda = \epsilon + i v k + \tfrac12 (1-v^2)k^2 + O(\epsilon k,k^3),
$$
so the logarithmic singularity in $\Phi$ and the marginal derivative singularity
in $\Psi$ induces a delicate coupling between $\epsilon$ and $k$.
Solving the equation $D(k,z)=0$ for $z=z(k)$ yields, for $k\to 0$,
\begin{equation}\label{eq:z-pole-critical}
1 - z(k)
\;\sim\;
i v k
\;+\;
C_1\,\frac{k^2}{|\log k|}
\qquad (k\to 0),
\end{equation}
with $C_1>0$ depending on $p,r,b$.  Thus the pole is dominated by the ballistic
drift term $ivk$, while the subleading correction---which controls the growth of
fluctuations---is the marginally suppressed quantity $k^2/|\log k|$.
Near the pole, the propagator behaves as:
$$
\tilde P(k,z)
\sim
\frac{\text{const}}{1 - z/z(k)},
\qquad
z\to z(k),
$$
and classical Tauberian theorems then convert the expansion (\ref{eq:z-pole-critical}) into asymptotic statements 
for the moments. The linear term $i v k$ enforces ballistic transport,
$$
\langle x_T\rangle \sim -\,v\,T,
\qquad v=2p-1,
$$
while the $k^2/|\log k|$ correction produces an ``almost ballistic'' growth of fluctuations:
$$
\mathrm{Var}(x_T)
\;\sim\;
C_2\,\frac{T^2}{\log T},
\qquad T\to\infty,
$$
with $C_2>0$.  Restoring the ballistic term in the second moment, we obtain:
\begin{align}
\langle x_T\rangle
&\sim -(2p-1)\,T, \\[4pt]
\langle x_T^2\rangle
&\sim (2p-1)^2 T^2
\;+\;
C_2\,\frac{T^2}{\log T}, \\[4pt]
\mathrm{Var}(x_T)
&\sim C_2\,\frac{T^2}{\log T}.
\end{align}
The dimensionless constant $C_2 = C_2(p,r,b)$ arises from the coefficients of
the logarithmic terms in $\Phi$, $\Psi$ and from the detailed structure of the
denominator $D(k,z)$.
\subsection{Regime $a>1$: Finite Mean Upstream Dwell (Supercritical)}
For $a>1$, the upstream dwell has finite mean $\tau_-=b/(a-1)$. The downstream dwell has finite mean $\tau_+=1/r$. 
Therefore, the process admits a long-time stationary fraction of time spent in each phase, and 
we expect a well-defined effective drift and, in principle, normal fluctuations.
The master-equation/generating-function analysis realizes this through the fact that $\Phi(\Lambda)$ and $\Psi(\Lambda)$ are finite at $\Lambda=1$ and 
analytic in a neighborhood. At $k=0$,
$$
H_+(0) = H_-(0) = 1,\quad \Lambda = z.
$$
The denominator of $\tilde P(k,z)$ becomes:
$$
D(0,z) = 1 - z\big[(1-r)+r\Psi(z)\big].
$$
The dominant pole in $z$ at $k=0$ is the root of $D(0,z)=0$. Physically, for a normalizable process starting at $t=0$, this root 
is at $z=1$. Consistency of probability conservation implies that:
$$
(1-r)+r\Psi(1) = 1,
$$
or $\Psi(1)=1$. This identity is simply the requirement of normalization of the generating function kernel.
To obtain the effective drift, we expand the root $z(k)$ of $D(k,z(k))=0$
for small $k$, and read off the first derivative at $k=0$. That is, we look for an expansion:
$$
z(k) = 1 + i \alpha_1 k - \alpha_2 k^2 + O(k^3),
$$
and the coefficient $\alpha_1$ is directly proportional to the effective velocity. Carrying out the expansion of $D(k,z)$ in
small $k$ and small $(1-z)$ (using that $\Phi$, $\Psi$ are smooth at $\Lambda=1$ for $a>1$), one 
finds that the first-order term in $k$ matches the time-averaged drift:
\begin{equation} \label{v(a)}
v(a) = (2p-1)\frac{(a-1)-rb}{(a-1)+rb}.
\end{equation}
Thus the mean displacement behaves ballistically:
$$
\langle x_T\rangle \sim v(a)\,T,\qquad T\to\infty,\ a>1.
$$
This formula can be intuitively understood by noting that the stationary fraction of time spent downstream is $\tau_+ /(\tau_++\tau_-)$, and 
upstream $\tau_- /(\tau_++\tau_-)$, so the net average sign of the phase is
$$
\langle\sigma\rangle
= \frac{\tau_+ - \tau_-}{\tau_+ + \tau_-}
= \frac{1/r - b/(a-1)}{1/r + b/(a-1)}
= \frac{(a-1) - rb}{(a-1) + rb}.
$$
Since $\mathbb{E}[Y_t\mid\sigma_t]=v\sigma_t$, we have $\mathbb{E}[Y_t]=v\,\langle\sigma\rangle$, giving (\ref{v(a)}).
To get $\langle x_T^2\rangle$, we examine the second derivative of $\tilde P$ in $k$ near the dominant pole $z(k)$. 
Technically, we rewrite $\tilde P(k,z)$ close to $D(k,z)=0$ as
$$
\tilde P(k,z) \approx \frac{N(k,z)}{D(k,z)},
$$
where $N(k,z)=1+r\Phi(\Lambda)$ is regular. The dominant contribution as $T\to\infty$ comes from the pole in $z(k)$, so the 
behavior of $\langle x_T\rangle$ and $\langle x_T^2\rangle$ is governed by the small-$k$ expansion of $z(k)$.
After a somewhat lengthy but standard expansion (keeping terms up to order $k^2$ in the pole position and in the residue), one finds that
the leading behavior of $\langle x_T^2\rangle$ is:
$$
  \langle x_T^2\rangle \sim v(a)^2 T^2 + D(a,b,r,p)T, \quad T\to\infty,
$$
 where $D(a,b,r,p)>0$ is an effective diffusion constant that can be expressed in terms of $H_\pm$ and the 
 values and derivatives of $\Phi,\Psi$ at $\Lambda=1$. The full expression for $D(a,b,r,p)$ is lengthy and not 
 particularly illuminating; it is a smooth, positive function of the parameters and exhibits no singular behavior for $a>1$.
Consequently,
$$
\mathrm{Var}(x_T)=\langle x_T^2\rangle-\langle x_T\rangle^2 \sim D(a,b,r,p)\,T.
$$

\begin{figure}[t]
\centering
\includegraphics[width=0.65\textwidth]{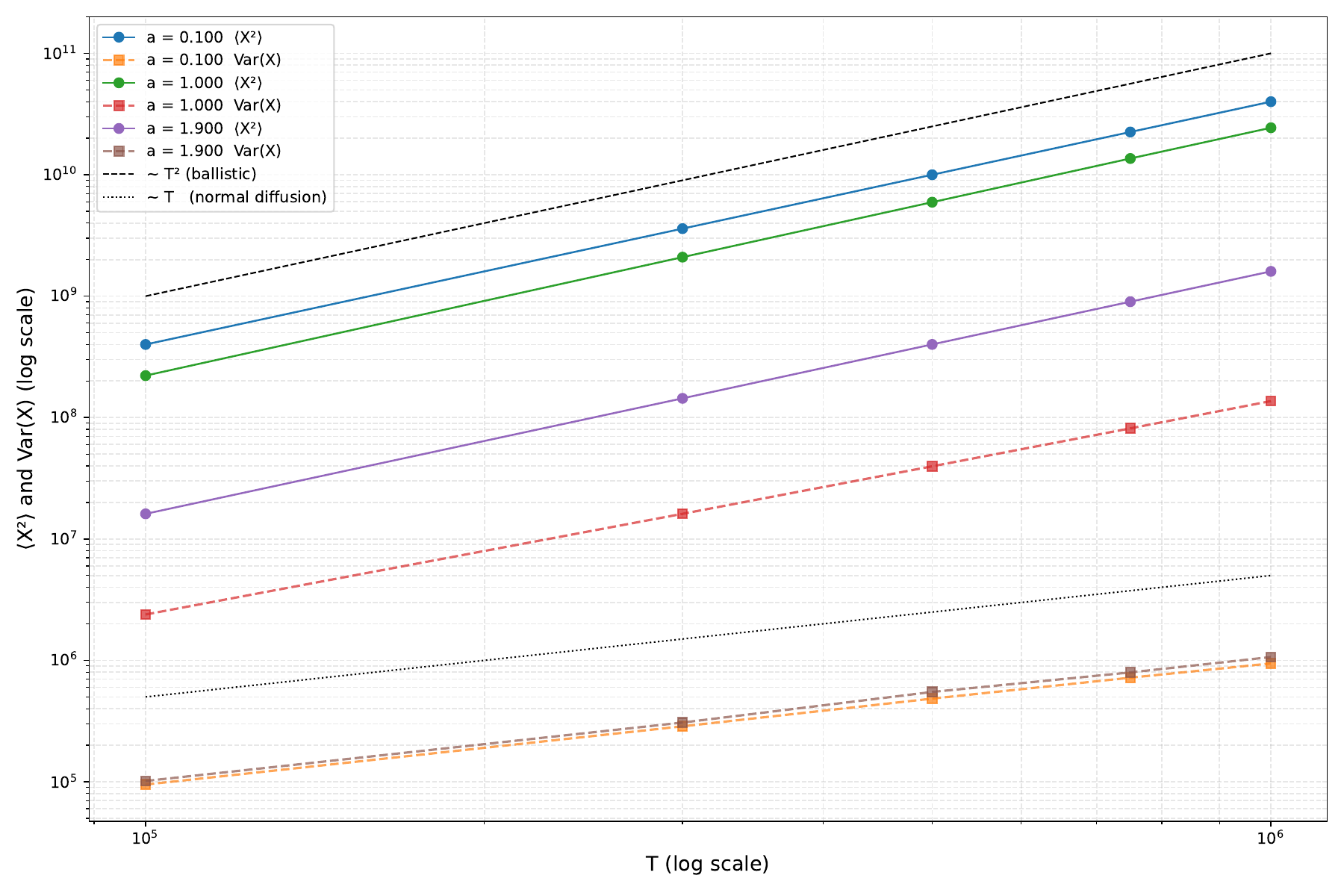} 
\caption{Scaling of the second moment $\langle x_T^2 \rangle$ (solid symbols) and variance $\mathrm{Var}(x_T) $ (open symbols) for a 
broad range of the exponent $a$. In the subcritical ($a=0.100$) and supercritical ($a=1.900$) regimes, the variance follows the 
ordinary diffusive scaling $\sim T$ (dotted line). At the critical point $a=1.000$, the variance exhibits a marked departure from 
normal diffusion, growing nearly ballistically.}
\label{fig:wide_range}
\end{figure}
\section{Numerical Validation via Monte Carlo Simulations}\label{s7}
To validate the analytical results derived in the preceding sections, we have performed Monte Carlo simulations of the persistent random walk 
defined in Section \ref{s2}. The simulations track the evolution of the first two moments, $\langle x_T^2 \rangle$ and $\mathrm{Var}(x_T) $, 
across the three distinct parameter regimes of $a$. Numerical data were collected for total observation times up to $T = 10^6$ steps, averaged over $10^4$ 
independent realizations to ensure statistical convergence. The fixed parameters are $p=0.6$, $r=0.6$ and $b=1$.

\begin{figure}[t]
\centering
\includegraphics[width=0.65\textwidth]{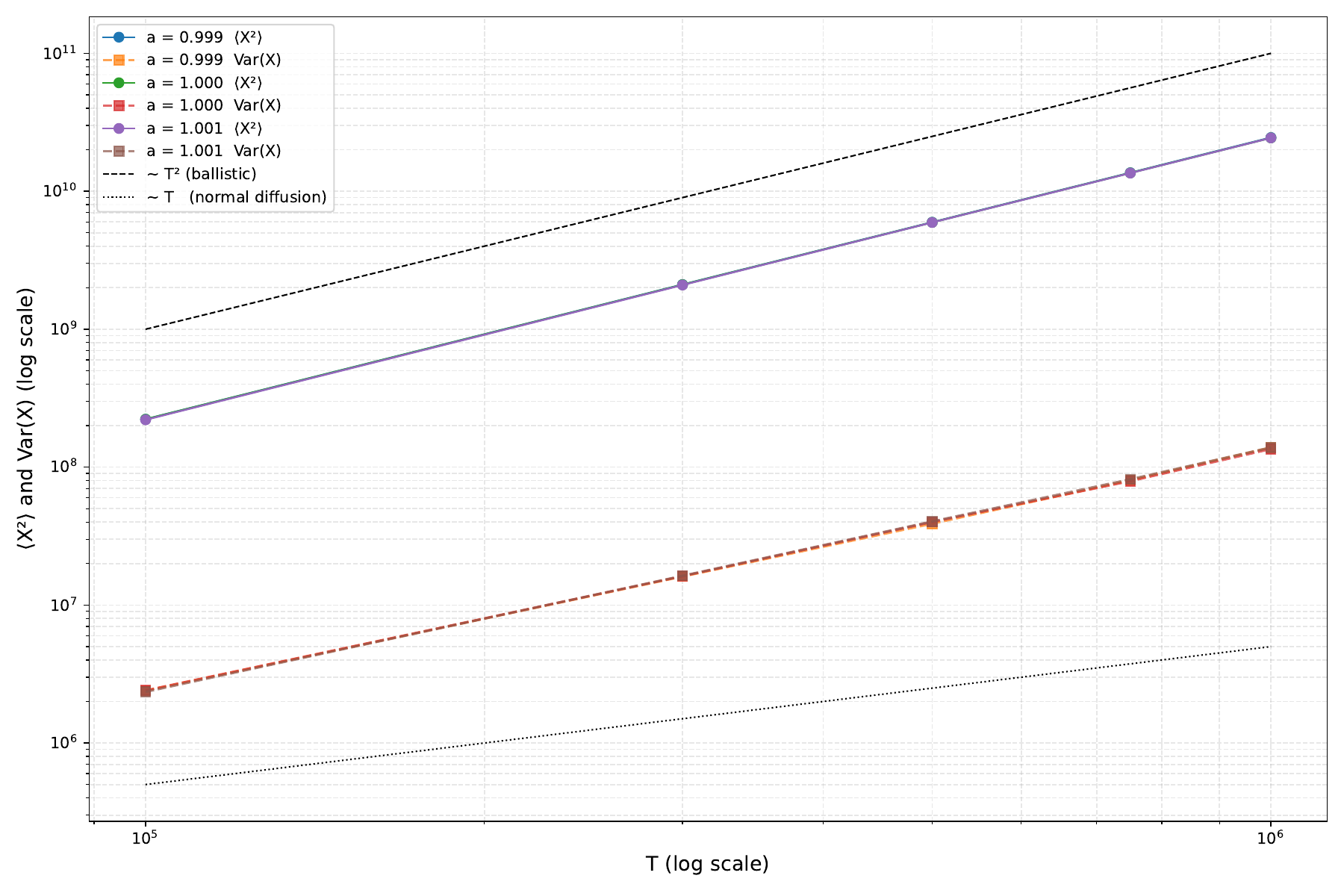} 
\caption{Detailed view of the scaling behavior in the vicinity of the critical point $a=1$. 
The variance for $a=1.000$ (red squares) clearly deviates from the diffusive 
$\sim T$ scaling and exhibits a near-ballistic growth that is systematically 
weaker than $\sim T^2$, in agreement with the marginal prediction derived 
in Section~\ref{s6}. For small deviations from criticality 
($a=0.999$ and $a=1.001$), the variance remains practically indistinguishable 
from the $a=1$ curve over the entire simulated time window, reflecting the 
extremely slow crossover away from the marginal scaling.}
\label{fig:critical_zoom}
\end{figure}

The simulation results, presented in Fig.~\ref{fig:wide_range} and Fig.~\ref{fig:critical_zoom}, demonstrate excellent agreement with the 
theoretical predictions. As established in the analytical framework, the second moment $\langle x_T^2 \rangle$ exhibits ballistic scaling 
($\sim T^2$) across all regimes, driven by the persistent bias $v$. The central interest, however, lies in the scaling of the fluctuations, 
$\mathrm{Var}(x_T)$, which undergoes a sharp transition at $a=1$.

In the subcritical regime ($a < 1$, blue curves), the walker is effectively localized in the upstream phase, leading to a persistent ballistic 
drift and ordinary diffusive fluctuations around the biased front. Similarly, in the supercritical regime ($a > 1$, purple curves), the 
finite mean upstream dwell time allows the system to reach a stationary phase mixture. Despite the heavy-tailed nature of the upstream 
dwell times for $1 < a < 2$, the simulations confirm the analytical prediction in Section \ref{s6}: the variance scales linearly with time 
($\mathrm{Var}(x_T) \propto T$), effectively matching the slope of the normal diffusion reference line.

The singular behavior at the marginal case $a=1$ is highlighted in 
Fig.~\ref{fig:critical_zoom}. The numerical data show that while the mean-square 
displacement remains close to ballistic scaling, the fluctuations are 
significantly enhanced compared to the supercritical regime. As derived in 
Section~\ref{s6}, the variance at $a=1$ follows an almost ballistic scaling, 
$\mathrm{Var}(x_T) \sim T^2 / \log T$. In a log--log representation this appears as a 
trajectory that is substantially steeper than the diffusive $\sim T$ line but slightly 
shallower than the ballistic $\sim T^2$ guide. The simulation data for $a=1.000$ (red squares) 
follow this anomalous scaling with high accuracy.

\section{Concluding Remarks}\label{s8}
In this work, we have provided a rigorous characterization of the asymptotic behavior of a discrete-time persistent random 
walk subject to an age-dependent upstream hazard. By employing a unified framework based on the master-equation and 
generating-function (propagator) method, we derived an exact expression for the Fourier--$z$ propagator $\tilde{P}(k, z)$. 
This approach allowed us to encapsulate the complex aging dynamics of the upstream phase into two fundamental kernels, 
$\Phi(\Lambda)$ and $\Psi(\Lambda)$.

The asymptotic behavior observed at the critical point $a = 1$ resembles yet fundamentally differs
from, L\'evy walk dynamics. In a standard unbiased L\'evy walk, the flight-time distribution exhibits a power-law 
tail $\psi(\tau) \sim \tau^{-(1+\alpha)}$ with $1 < \alpha < 2$, while the walker moves at constant speed between 
turning events. This leads to superdiffusive growth of the mean squared displacement,
$\langle x_T^2 \rangle \sim T^{3-\alpha}$ \cite{Zaburdaev2015}. In our model, the upstream run durations (periods of 
persistent motion against the flow) follow a power-law distribution $\psi(\tau) \sim \tau^{-(1+a)}$, originating 
from the age-dependent hazard $h_c = a/(b+c)$. When $a < 1$, the mean upstream run time diverges, yet the variance 
of the displacement remains diffusive, $\mathrm{Var}(x_T) \sim T$, rather than superdiffusive. 

This qualitative difference stems from the presence of a finite drift in our dynamics. Whereas unbiased L\'evy walks 
have zero net drift, our walker possesses a ballistic bias $v = 2p-1$ in both phases. As a consequence, the heavy-tailed 
run statistics do not translate into superdiffusive spreading but instead modify the fluctuations around the dominant 
drift. The bias effectively masks the influence of the heavy-tailed excursions at the level of the variance, leading 
to ordinary diffusion for $a>1$ and to the marginal scaling $\mathrm{Var}(x_T) \sim T^2/\log T$ at the critical point 
$a=1$. 

Our process therefore occupies an intermediate position between unbiased L\'evy walks, where superdiffusion arises 
from the coupling of space and time in the absence of drift \cite{Klafter2011}, and biased persistent random walks with exponential 
switching times. The point $a=1$ marks a marginal breakdown of the Central Limit Theorem, where the logarithmic 
correction to ballistic scaling signals the onset of anomalously large fluctuations.

It is instructive to connect our results to the large-deviation analysis of the same model in \cite{SJ2026}. There, the system was studied 
within the $s$-ensemble, where a counting field biases the time-integrated current. A dynamical phase transition was found at a critical 
value of the bias parameter, corresponding to a non-analyticity in the scaled cumulant generating function. In the present work, we focus 
on the unbiased ensemble ($s=0$) and examine the bulk moments of the displacement. The two approaches are complementary: the 
large-deviation framework characterizes the exponential suppression of current fluctuations, while our generating-function analysis provides 
the exact scaling of the moments themselves. The singular behavior of the large-deviation rate function is directly reflected 
in the growth law of the second cumulant.

Several avenues for future research remain open. First, the current model assumes a constant hazard probability for the 
downstream phase; extending this to a symmetric age-dependent hazard for both phases could reveal more complex 
competitive aging dynamics. Second, the introduction of an external spatial potential or quenched disorder could 
significantly alter the localization-to-transport transition observed here. Finally, the continuous-time limit of this discrete 
model could provide further insights into the relationship between age-dependent hazards and fractional diffusion 
equations. Such extensions would be particularly relevant for modeling biological foraging strategies and transport in 
disordered media, where intermittent search patterns and aging effects are ubiquitous.
 
\end{document}